\documentclass[12pt]{iopart}

\usepackage{iopams}  
\usepackage{cite}

\usepackage{graphicx}
\usepackage[usenames]{color}

\begin{document}

\title{Perron-Frobenius theorem on the superfluid transition of an ultracold Fermi gas}
\author{Naoyuki~Sakumichi$^1$, Norio~Kawakami$^1$ and Masahito~Ueda$^{2,3}$}
\address{$^1$Department of Physics, Kyoto University, Kyoto 606-8502, Japan}
\address{$^2$Department of Physics, The University of Tokyo, 7-3-1 Hongo, Tokyo 113-0033, Japan}
\address{$^3$ERATO Macroscopic Quantum Project, JST, Tokyo 113-0033, Japan}

\ead{sakumichi@scphys.kyoto-u.ac.jp}


\begin{abstract}
The Perron-Frobenius theorem is applied to identify the superfluid transition of a two-component Fermi gas with a zero-range s-wave interaction.
According to the quantum cluster expansion method of Lee and Yang,
the grand partition function is expressed by the Lee-Yang contracted 0-graphs.
A singularity of an infinite series of ladder-type Lee-Yang contracted 0-graphs is analyzed.
We point out that the singularity is governed by the Perron-Frobenius eigenvalue of a certain primitive matrix which is defined in terms of the two-body cluster functions and the Fermi distribution functions.
As a consequence, it is found that there exists a unique fugacity at the phase transition point, which implies that there is no fragmentation of Bose-Einstein condensates of dimers and Cooper pairs at the ladder-approximation level of Lee-Yang contracted 0-graphs.
An application to a Bose-Einstein condensate of strongly bounded dimers is also made.
\end{abstract}

\submitto{\NJP}


\maketitle

\section{INTRODUCTION}

When physically relevant operators are primitive in some representation, the Perron-Frobenius theorem may play a role in the study of physics.
Examples include theories of stochastic processes \cite{S80} and studies of ground states of quantum many-body systems \cite{LM62, T89}.
In this study, we provide a specific example of primitive matrices,
which belongs to a different category.
It concerns a two-body cluster function in quantum-statistical mechanics, which is important to dilute Bose and Fermi systems.

To be specific, we consider two-component atomic Fermi gases with a zero-range interaction characterized by the s-wave scattering length $a$.
This system exhibits a crossover between a BCS-like state of Cooper pairs and a Bose-Einstein condensate (BEC) of diatomic molecules \cite{RZZ12}.
The two-body cluster function $U^{(2)}$ mentioned above appears in the quantum cluster expansion of the grand partition function.
The quantum cluster expansion is important to investigate thermodynamic properties of the system, and it has seen a resurgence of interest recently \cite{HM04, R07, LHD09, LHD10, KS11,DB12}.
The cluster integrals $b^{(3)}$ and $b^{(4)}$ for the unitary Fermi gas have been experimentally measured using a mixture of $^6$Li in two internal states \cite{NNJCS10, KSCZ11}.
Here the cluster expansion is
$\beta p \lambda^3 = 2 + \frac{3\sqrt{2}}{4} z^2 + b^{(3)} z^3 + b^{(4)} z^4 +\cdots$, where $p$ is the pressure and 
 $\lambda := (2\pi /m k_{\rm B} T)^{1/2}$ is the thermal de Broglie length.
The second cluster integral $b^{(2)}=\frac{3\sqrt{2}}{4}$ can be calculated from the trace of the two-body cluster function $U^{(2)}$.
In this regard, the two-body cluster function $U^{(2)}$ has more information than the second cluster integral $b^{(2)}$.
Under the s-wave approximation, we show that $U^{(2)}$ is a non-negative matrix with the Bethe-Peierls boundary condition \cite{BP35} or the Fermi pseudopotential \cite{F36}.

According to the quantum cluster expansion method of Lee and Yang \cite{LY58-I, LY60-IV},
the grand partition function is calculated by the cluster functions $U^{(1)}$, $U^{(2)}$, etc.,
and expressed by the Lee-Yang contracted 0-graphs.
In this study, we consider an infinite series of the ladder-type Lee-Yang contracted 0-graphs,
which gives the non-interacting dimer BEC transition temperature in BEC limit \cite{OU06, SKU11}, and the BCS-transition temperature in BCS limit \cite{future}.
To investigate the phase transition of the system,
we analyze a singularity of an infinite series of ladder-type Lee-Yang contracted 0-graphs.
We point out that the singularity is governed by the Perron-Frobenius eigenvalue of a certain primitive matrix which is defined in terms of the two-body cluster functions $U^{(2)}$ and the Fermi distribution functions $n_F$.
As a consequence, there exists one and only one fugacity at a phase transition point and physically it implies that there is no fragmentation 
of Bose-Einstein condensates of dimers and Cooper pairs at the ladder-approximation level of the Lee-Yang contracted $0$-graphs.

This paper is organized as follows.
In section 2, the phase transition is defined as the disappearance of holomorphy in thermodynamic functions.
In section 3, the Lee-Yang cluster expansion method is reviewed for the case of two-component Fermi systems.
In section 4, the matrix elements of the two-body cluster functions in momentum representation are calculated and are found to be non-negative.
In section 5, a singularity of an infinite series of the ladder-type Lee-Yang contracted graphs are analyzed by using the Perron-Frobenius theorem.
An application to the BEC of strongly bounded dimers is also made.
In section 6, we summarize the main results of this paper.
The proofs of several formulas and inequalities are relegated to appendices to avoid digressing from the main subject.

\section{Grand partition function and phase transition}

To be specific, we consider a system of two-component Fermi particles with the same mass $m$ in a finite volume $V=L^3$ at a temperature $T$.
We also introduce the fugacity $z=\rme^{\beta \mu}$, where $\beta=1/k_B T$ is the inverse temperature and $\mu$ is the chemical potential per particle which is assumed to be independent of spin variables $\sigma = \uparrow, \downarrow$. 
Our $N$-particle Hamiltonian and the grand partition function are given by
\begin{eqnarray}
& H^{(N)} = - \frac{1}{2m}
\sum_{i=1}^N \nabla_i^2 + \sum_{i<j} v\left(\left|\bold{r}_i-\bold{r}_j \right| \right), \\
%
& \Xi_V
  := 1 +  \sum_{N=1}^\infty  z^N \Tr_{\mathcal{H}_{\rm{A}}^{(N)}} 
  \left[  \rme^{-\beta H^{(N)}}\right],
\end{eqnarray}
where $\Tr_{\mathcal{H}_{\rm{A}}^{(N)}} \! \left[ \cdots \right]$ denotes
the trace over the antisymmetrized $N$-particle Hilbert space $\mathcal{H}_{\rm{A}}^{(N)}$.
Here we set $\hbar = 1$.
The equilibrium pressure $p$ and the particle-number density $\rho$ of the system are given by
\begin{equation}
p  = \lim_{V\to\infty} \frac{1}{\beta V} \log \Xi_V,
\label{eq:pressure}
\end{equation}
and
\begin{equation}
 \rho = \lim_{V\to\infty}
    \frac{1}{V} \, z \, \frac{\partial}{\partial z}
    \log \Xi_V.
\label{eq:densitiy}
\end{equation}

A phase transition is mathematically characterized 
by singularities in thermodynamic functions.
The singularities here imply the disappearance of holomorphy.
A complex function defined on $\mathbb{C}$ is holomorphic on an open set $\Omega$, 
if its Taylor expansion around every point in $\Omega$ has a nonzero radius of convergence. 
If $\Omega$ is not an open set we interpret that holomorphy holds in an appropriate open set containing $\Omega$. 
In literature, the term `analytic' is often used instead of holomorphic; however, we use the latter in this paper, because the former is a global concept.
In the classical limit,
the cluster expansion of $p$ has a positive radius of convergence \cite{LP66},
so if the fugacity $z$ is sufficiently small, $p$ is holomorphic in $z$. 
That is, there is no phase transition if $z \ll 1$.
If a singularity in a thermodynamic function such as $p(z)$ exists at a point $z_0$ on the positive real axis
and the holomorphy disappears there,
we conclude that there is a phase transition.
Lee and Yang analyzed zeros of $\Xi$ for classical lattice gas systems (or equivalently classical Ising systems \cite{YL52-I}.
In contrast, for the superfluid phase transition, we analyze zeros of $\Xi^{-1}$.

Now we assume that a binary interaction $v$ is characterized by the s-wave scattering length $a$ and that all other relevant length scales, such as the thermal de Broglie length $\lambda := \sqrt{2\pi \hbar^2 /mT}$ and the mean interparticle distance, are much larger than the range of interaction.
Then, by dimensional analysis, we have
\begin{equation}
 \lim_{V \to \infty} \frac{1}{V} \log \Xi_V
 = \frac{1}{\lambda^3} f\left( z,  \lambda/a  \right).
\end{equation}
The superfluid transition manifests itself through the singularity of $f(z,\lambda/a)$ on the positive real axis of $z$.

\section{Lee-Yang cluster expansion method for a two-component Fermi system}

In the Lee-Yang cluster expansion method, 
the grand partition function for a system of Fermi particles can be expressed in terms of the contracted $0$-graphs \cite{LY60-IV}.
Each contracted $0$-graphs is computed from the Fermi distribution function 
$ n_{\rm F}( \bold{k} )
 :=   ( 1+ z^{-1} \rme^{\beta \bold{k}^2/(2m)} )^{-1}$
and the $N$-particle cluster functions $U^{(N)}$ for the same system obeying Boltzmann statistics.

In this section, we first define the $N$-particle cluster functions $U^{(N)}$.
Then, we show how to express the grand partition function in terms of the contracted $0$-graphs.

\subsection{Definition of Cluster functions}

To define the $N$-particle cluster function, we first introduce
\begin{equation}
\fl
 \langle 1,\dots , N | W^{(N)}  | 1,\dots , N \rangle 
  := N ! \!\!\!\! \sum_{\,\,\,\, \psi_i \in \mathcal{H}^{(N)}} \!\!\!
 \psi_i \left( 1', \dots , N' \right)
  \psi_i^* \left( 1,\dots , N \right)  \rme^{-\beta E_i},
\label{eq:def:W}
\end{equation}
where we use $1,2,\dots$ and $1',2',\dots$ 
to represent the momentum and spin coordinates of individual, $1 := (\bold{k}_1,\sigma_1)$, etc. and $1' := (\bold{k}_1',\sigma_1')$, etc.,
and $\psi_i$ and $E_i$ are the normalized eigenfunctions and eigenvalues of 
 $H^{(N)}$, respectively.
The summation in Eq.~(\ref{eq:def:W}) extends over all eigenfunctions $\psi_i$ in an unsymmetrized $N$-particle Hilbert space $\mathcal{H}^{(N)}$.
The matrix elements of $U^{(N)}$ in momentum representation are
\begin{eqnarray}
    & \langle 1' | U^{(1)} | 1 \rangle
     := \langle 1' | \rme^{-\beta H^{(1)}} | 1 \rangle, \\
    & \langle 1', 2' | U^{(2)} | 1, 2 \rangle
     :=  \langle 1', 2' | \rme^{-\beta H^{(2)}} | 1, 2 \rangle
      -  \langle 1' | U^{(1)} | 1 \rangle \, \langle 2' | U^{(1)} | 2 \rangle,\\
     & \fl \langle 1', 2' , 3' | U^{(3)} |  1, 2, 3 \rangle 
      :=\langle 1', 2' , 3' | \rme^{-\beta H^{(3)}} | 1, 2 , 3 \rangle 
      -  \langle 1' |  U^{(1)} | 1 \rangle \langle 2', 3' | U^{(2)} | 2, 3 \rangle \nonumber \\
     & \qquad - \langle 2' |  U^{(1)} | 2 \rangle \langle 3', 1' | U^{(2)} | 3, 1 \rangle 
      - \langle 3' |  U^{(1)} | 3 \rangle \langle 1', 2' | U^{(2)} | 1, 2 \rangle \\
      & \qquad -\langle 1' |  U^{(1)} | 1 \rangle \langle 2' | U^{(1)} | 2 \rangle \langle 3' | U^{(1)} | 3 \rangle
        ,\quad {\rm etc.}, \nonumber
\label{eq:Ursell}
\end{eqnarray}
In the computation of the grand partition function for Fermi particles, only the anti-symmetric combination $\Upsilon_{\rm A}^{(N)}$ appears.
We define the matrix elements of $\Upsilon^{(N)}_{\rm{A}}$ by
\begin{equation}
\fl \bigl\langle 1',\dots ,l'\, \big| 
     \Upsilon^{(N)}_{\rm{A}} 
  \big|\, 1,\dots ,l\, \bigr\rangle 
 :=  \sum_{P \in S_N} (-1)^P
\bigl\langle P(1'),\dots ,P(l')\, \big| 
     U^{(N)}
  \big|\, 1,\dots ,l\, \bigr\rangle,
  \label{eq:Upsilon_A}
\end{equation}
where $P$ denotes permutation and $(-1)^P=1$ or $-1$ for even or odd permutations $P$.

We discuss a couple of examples.\\

\noindent
{\it Example 1}---

The one-particle cluster function is
\begin{equation}
    \langle \bold{k}' \sigma'| U^{(1)}  | \bold{k}\sigma \rangle 
   = \delta_{\sigma,\sigma'} \langle \bold{k}' | U^{(1)}  | \bold{k} \rangle   
   = \delta_{\bold{k},\bold{k}'} \delta_{\sigma,\sigma'}\rme^{-\beta \bold{k}^2/(2m)}.
\label{eq:U1}
\end{equation}
We consider the geometric series 
\begin{equation}
   \sum_{n=0}^{\infty} \left(- z \, \rme^{-\beta \bold{k}^2/(2m)} \right)^n
   = \left( 1 + z \, \rme^{-\beta \bold{k}^2/(2m)} \right)^{-1}
   = 1 - n_{\rm F}( \bold{k} ).
\end{equation}
The effect of Fermi statistics emerges through this geometric series.\\

\noindent
{\it Example 2}---

The momentum part of the two-body cluster function is given by
\begin{eqnarray}
    \langle \bold{k}_1' \sigma_1',\bold{k}_2' \sigma_2'| U^{(2)}  | \bold{k}_1 \sigma_1,\bold{k}_2 \sigma_2 \rangle 
   = \delta_{\sigma_1\sigma_1'} \delta_{\sigma_2\sigma_2' }
   \langle \bold{k}_1' ,\bold{k}_2' | U^{(2)}  | \bold{k}_1 ,\bold{k}_2  \rangle .
\end{eqnarray}
In this paper, we discuss the case of two-component Fermi particles with an s-wave approximation.
Then, a two-body cluster function is a symmetry matrix
$
 \langle \bold{k}_1', \bold{k}_2' | U^{(2)} | \bold{k}_1,\bold{k}_2 \rangle 
 = \langle \bold{k}_2', \bold{k}_1' | U^{(2)} | \bold{k}_1,\bold{k}_2 \rangle
$, 
and
\begin{eqnarray}
 \langle 1',2' | \Upsilon_{\rm{A}}^{(2)} | 1,2 \rangle 
 = \langle \bold{k}_1', \bold{k}_2' | U^{(2)} | \bold{k}_1,\bold{k}_2 \rangle 
 \cdot \left( \delta_{\sigma_1\sigma_1'} \delta_{\sigma_2\sigma_2'}
    - \delta_{\sigma_1\sigma_2'}\delta_{\sigma_2\sigma_1'}   \right).
\end{eqnarray}
The two-particle Hamiltonian is
\begin{equation}
H^{(2)}= \frac{1}{2m} \nabla_1^2 + \frac{1}{2m} \nabla_2^2 + v (\bold{r}) =\frac{1}{4m}\nabla_{\bold{R}}^2+H^{({\rm rel})},
\end{equation}
\begin{equation}
H^{({\rm rel})} = \frac{1}{m}\nabla_{\bold{r}}^2 + v  (\bold{r}).
\end{equation}
The two-body cluster function for a finite volume $V$ is
\begin{eqnarray}
 \langle\bold{k}_1'\bold{k}_2'|U^{(2)}|\bold{k}_1\bold{k}_2\rangle  
 = \frac{8\pi^3}{V} \delta_{\bold{K},\bold{K}'} \,
     \rme^{- \beta\bold{K}^2/(4m)} 
     \cdot \langle\bold{k}'| u^{({\rm rel})}  |\bold{k}\rangle,
\label{eq:U2}
\end{eqnarray}
\begin{equation}
\langle\bold{k}'| u^{({\rm rel})}  |\bold{k}\rangle
:= \langle\bold{k}'| \rme^{-\beta H^{({\rm rel})}}  |\bold{k}\rangle -  \delta_{\bold{k},\bold{k}'} \, \rme^{-\beta \bold{k}^2/m},
\label{eq:Urel} 
\end{equation}
where $\bold{K}:=\bold{k}_1+\bold{k}_2$ and $\bold{k}:=(\bold{k}_1-\bold{k}_2)/2$, and the Kronecker delta reflects the conservation of momentum.
The function $\langle\bold{k}'| u^{({\rm rel})}  |\bold{k}\rangle$ describes an effect of interaction and can be calculated from the eigenfunctions and eigenvalues of $H^{({\rm rel})}$ \cite{LY58-I}.

\subsection{The grand partition function represented by contracted $0$-graphs}

\begin{figure}
\begin{picture}(220,135)
\put(75,124){(a)}
\put(97,65){\includegraphics[width=330pt]{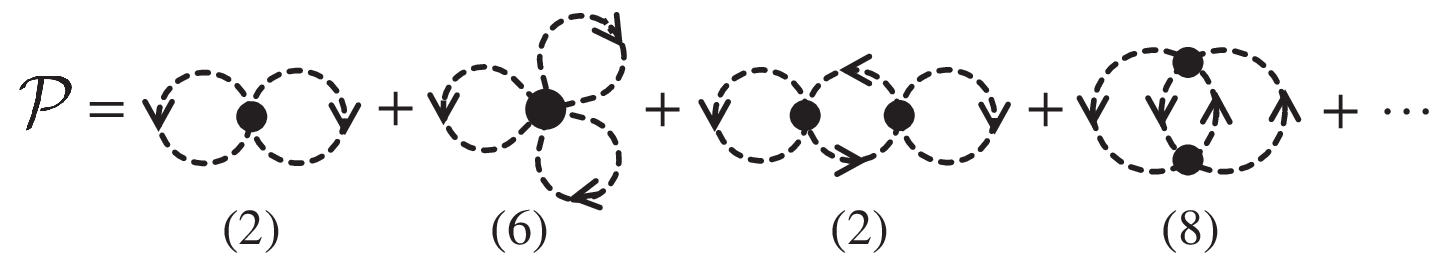}}
\put(75,52){(b)}
\put(100,-5){\includegraphics[width=225pt]{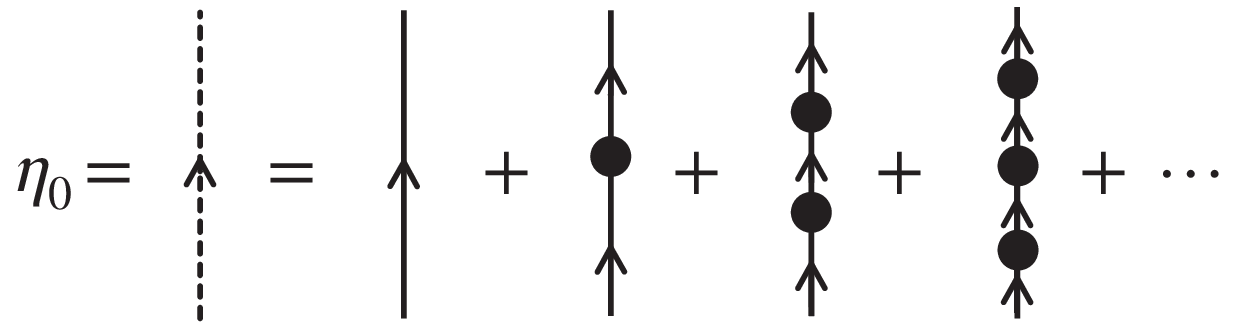}}
\end{picture}
\caption{
(a) Sum over all different contracted $0$-graphs (dotted curves).
The symmetry numbers are shown under the graphs.
(b) Expression of
$\langle \bold{k}'\sigma' | \eta_0 | \bold{k}\sigma \rangle$
as the sum over different primary $1$-graphs (solid lines).}
\label{fig:Xi}
\end{figure}

In terms of the cluster functions,
the grand partition function $\Xi$ is written as \cite{LY58-I, LY60-IV}
\begin{equation}
\log \Xi_V =
- 2 \frac{V}{\lambda^3} {\rm Li}_{\frac{5}{2}} (-z) 
+ \mathcal{P},
\label{eq:logX-contracted}
\end{equation}
where ${\rm Li}_l(x):=\sum_{n=1}^\infty x^n/n^{l}$ is the polylogarithm and 
the first term on the right-hand side gives the logarithms of the grand partition function of a non-interacting Fermi gas.
Here $\mathcal{P}$ is the sum of all different contracted $0$-graphs introduced by Lee and Yang \cite{LY60-IV}
and is illustrated in figure~\ref{fig:Xi}.
The number under each graph in figure~\ref{fig:Xi} shows the symmetry number of the corresponding contracted $0$-graph.
The algebraic expression of the sum of the contracted graphs is
\begin{eqnarray}
& \fl \mathcal{P}
 =  \frac{z^2}{2}
 \sum_{\bold{k}_1,\bold{k}_{2}}
   \sum_{\sigma_1,\sigma_{2}}
     \eta_0 (\bold{k}_1) \eta_0 (\bold{k}_{2}) 
 \langle \bold{k}_1\sigma_1,\bold{k}_2\sigma_2 | \Upsilon_{\rm{A}}^{(2)} | \bold{k}_1\sigma_1,\bold{k}_2\sigma_2 \rangle \nonumber \\
& \fl + \frac{z^3}{6}
 \!\! \sum_{\bold{k}_1,\bold{k}_2,\bold{k}_3}
   \sum_{\sigma_1,\sigma_2,\sigma_3} \!\!\!
     \eta_0 (\bold{k}_1) \eta_0 (\bold{k}_{2})  \eta_0 (\bold{k}_{3})
 \langle \bold{k}_1\sigma_1,\bold{k}_2\sigma_2,\bold{k}_3\sigma_3 | \Upsilon_{\rm{A}}^{(3)} | \bold{k}_1\sigma_1,\bold{k}_2\sigma_2,\bold{k}_3\sigma_3 \rangle  \nonumber\\
& \fl +  \frac{z^{4}}{2}
 \!\! \sum_{\bold{k}_1,\dots,\bold{k}_{4}}
   \sum_{\sigma_1,\dots,\sigma_{4}} \!\!\!
    \eta_0 (\bold{k}_1) \cdots \eta_0 (\bold{k}_{4}) 
  \langle \bold{k}_1\sigma_1,\bold{k}_2\sigma_2 | \Upsilon_{\rm{A}}^{(2)} | \bold{k}_1\sigma_1,\bold{k}_3\sigma_3 \rangle  
     \langle \bold{k}_{3}\sigma_{3},\bold{k}_{4}\sigma_{4} | \Upsilon_{\rm{A}}^{(2)} | \bold{k}_2\sigma_2,\bold{k}_4\sigma_4\rangle  \nonumber \\
& \fl +  \frac{z^{4}}{8}
 \!\! \sum_{\bold{k}_1,\dots,\bold{k}_{4}}
   \sum_{\sigma_1,\dots,\sigma_{4}} \!\!\!
    \eta_0 (\bold{k}_1) \cdots \eta_0 (\bold{k}_{4}) 
  \langle \bold{k}_1\sigma_1,\bold{k}_2\sigma_2 | \Upsilon_{\rm{A}}^{(2)} | \bold{k}_3\sigma_3,\bold{k}_4\sigma_4 \rangle  
     \langle \bold{k}_{3}\sigma_{3},\bold{k}_{4}\sigma_{4} | \Upsilon_{\rm{A}}^{(2)} | \bold{k}_1\sigma_1,\bold{k}_2\sigma_2\rangle
\nonumber \\
& \fl + \cdots, \label{eq:XiLad-Irr}
\end{eqnarray}
where each term in the sum corresponds to the contracted $0$-graph at the corresponding order in figure~\ref{fig:Xi}.
Here,
\begin{eqnarray}
 \eta_{0}( \bold{k} ) := 1 - n_{\rm F}( \bold{k} )
 =   ( 1+ z\,\rme^{-\beta \bold{k}^2/(2m)} )^{-1}
 \end{eqnarray}
describes the effect of the Fermi statistics.

\section{Two-body cluster function in the s-wave approximation}

To calculate an analytical representation of
$\langle\bold{k}_1', \bold{k}_2'|U^{(2)}|\bold{k}_1, \bold{k}_2\rangle$
within an s-wave approximation,
we can use the Bethe-Peierls boundary condition \cite{BP35} 
\begin{equation}
\frac{1}{r \psi} \frac{d}{dr} (r \psi) \Big|_{r \to +0} = - \frac{1}{a},
\label{eq:B-P}
\end{equation}
or the Fermi pseudopotential \cite{F36}
\begin{equation}
 v_{\rm ps} (\bold{r})
 :=  \frac{4\pi a}{m} \delta^3(\bold{r}) \frac{\partial}{\partial r} r,
\label{eq:def:pseudopotential}
\end{equation}
where $a$ is the s-wave scattering length.
The Bethe-Peierls boundary condition or the Fermi pseudopotential supports continuous scattering states
\begin{equation}
\psi_{\rm sc}(r)=\left\{ 2\pi^2[1+(k_{\rm sc}a)^2] \right\}^{-\frac{1}{2}}
   \frac{1}{r}
    \bigl(  \sin (k_{\rm sc}r) -k_{\rm sc}a \cos (k_{\rm sc}r)  \bigr),
\label{eq:def:scat-w.f.}
\end{equation}
with the energy $E_{\rm sc} = k_{\rm sc}^2/m$,
and if $a^{-1}>0$ one bound state
\begin{equation}
\psi_{\rm b}(r)= (2\pi a)^{-\frac{1}{2}} \, \frac{1}{r} \, \rme^{-r/a},
\label{eq:def:bound-w.f.}
\end{equation}
with the binding energy $E_{\rm b} = - 1/(ma^2)$.
Using the set of energy eigenvalues and eigenstates (\ref{eq:def:scat-w.f.}) and (\ref{eq:def:bound-w.f.}),
the two-body cluster function can be computed as
\begin{equation}
 \langle\bold{k}_1', \bold{k}_2'|U_{\rm BP}^{(2)}|\bold{k}_1, \bold{k}_2\rangle
 = \frac{8\pi^3}{V} \delta_{\bold{K},\bold{K}'} \,
     \rme^{- \beta\bold{K}^2/2} 
     \cdot \langle\bold{k}'| u_{\rm BP}^{({\rm rel})}  |\bold{k}\rangle,
\label{eq:U2_pseudopotential_finiteV}
\end{equation}
where
\begin{equation}
  \langle\bold{k}'|u_{\rm BP}^{({\rm rel})}|\bold{k}\rangle 
  = \frac{\lambda^3}{2^{5/2}\pi^{7/2}} \, \frac{s(x', w) -  s(x, w)}{x'^2-x^2},
\label{eq:Urel_pseudo}
\end{equation}
\begin{equation}
s(x, w)
 =  \frac{1}{x^2+w^2}
  \left(
     w \, \rme^{-x^2}
     -  \frac{2}{\sqrt{\pi}} \, x F \left(x \right)      
     -  w \, \rme^{w^2} {\rm erfc}  \left( -w \right)
    \right).
\label{eq:def:sx}
\end{equation}
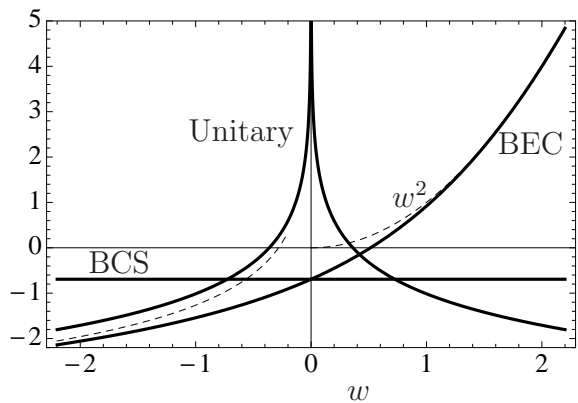
\begin{figure}
\centering
\includegraphics[width=215pt,clip]{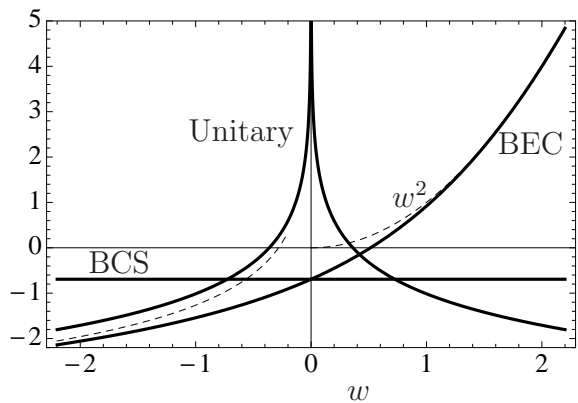}
\put(-87,-8){$w$}
\put(-147,90){Unitary}
\put(-30,85){BEC}
\put(-185,40){BCS}
\put(-70,65){$w^2$}
\caption{
The logarithms of three terms in the function $s(x, w)$.
``BCS", ``Unitary", and ``BEC" show
 $-\log 2$, $-\log(\left| w \right|)+0.3624$, 
and $w^2+\log(\frac{1}{2}{\rm erfc}(-w))$, respectively.
Here $0.3624\simeq{\rm Max} (xF(x)/\sqrt{\pi})$ ($x\simeq1.502$).
The dashed curves are $-\log(2\sqrt{\pi}\left| w \right|)$ and $w^2$.
}
\label{three-terms_of_gx}
\end{figure}
Here we introduce the dimensionless variables
$x:=\sqrt{\beta k^2/m}= \lambda k/\sqrt{2\pi}$  and $w:=\sqrt{\beta}/(\sqrt{m}a)=\lambda/(\sqrt{2\pi}a)$,
and the Dawson's integral 
\begin{equation}
F \left( x \right) =  \rme^{-x^2} \int_0^{x} dt \, \rme^{t^2},
\label{eq:def:DawsonF}
\end{equation}
 and the complementary error function
\begin{equation}
{\rm erfc}\left(x \right)
= \frac{2}{\sqrt{\pi}} \int_{x}^{\infty} \! dt \, \rme^{-t^2}.
\label{eq:def:erfc}
\end{equation}
The proof of (\ref{eq:U2_pseudopotential_finiteV})-(\ref{eq:def:sx}) is given in \ref{app:U2_pseudo}.
The two-body cluster function in the s-wave approximation has been first obtained in Ref.~\cite{OU06} for $a^{-1} \geq 0$ and in Ref.~\cite{In09} for $a^{-1} \! <0$.
However the representation in (\ref{eq:U2_pseudopotential_finiteV})-(\ref{eq:def:sx}) has two advantages over that in Ref. \cite{OU06, In09}
: (i) it holds for either sign of $a$
and (ii) it can easily be seen that the two-body cluster function is continuously connected at the unitary limit $a^{-1}=0$.
The function $s(x, w)$ in (\ref{eq:def:sx}) is composed of the three terms.
In the BCS region ($\lambda/a = \sqrt{2\pi\beta}/(\sqrt{m}a) < -1$),
the first term (i.e., $\rme^{-x^2}$) is dominant.
In the unitarity region ($\lambda/\left| a \right| < 1$),
the second term (i.e., $-  \frac{x}{w} \frac{2}{\sqrt{\pi}} F \left(x \right) $) is dominant.
In the BEC region ($\lambda/a > 1$),
the last term (i.e., $-  \rme^{w^2} {\rm erfc}  \left( -w \right)$) is dominant.
The logarithms of these three terms are shown in figure \ref{three-terms_of_gx}.

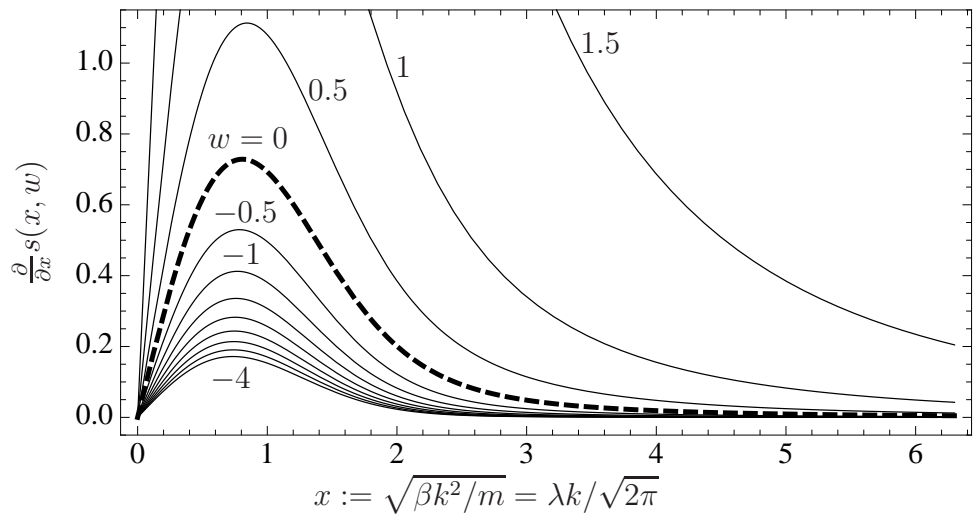
\begin{figure}
\centering
\includegraphics[width=340pt,clip]{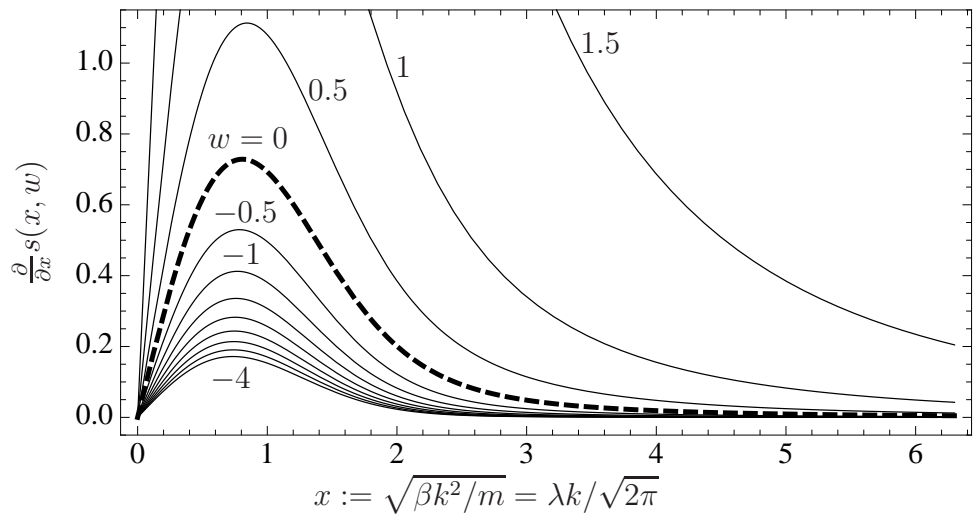}
\put(-250,-13){$x:=\sqrt{\beta k^2/m}= \lambda k/\sqrt{2\pi}$}
\put(-364,70){\rotatebox{90}{$\frac{\partial}{\partial x}s(x, w)$}}
\put(-289,31){$-4$}
\put(-285,78){$-1$}
\put(-288,94){$-0.5$}
\put(-290,122){$w=0$}
\put(-252,140){$0.5$}
\put(-219,149){$1$}
\put(-149,158){$1.5$}
\caption{Dependence of $\frac{\partial}{\partial x}s(x, w)$ on $x:=\sqrt{\beta k^2/m}= \lambda k/\sqrt{2\pi}$  for $w:=\sqrt{\beta}/(\sqrt{m}a)=\lambda/(\sqrt{2\pi}a) = -4,-3.5,-3,\dots,1,1.5$.
The dashed curve shows the case of the unitarity limit ($w = 0$).
We see that $\frac{\partial}{\partial x} s(x, w) >0$ for arbitrary $- \infty <w$ and $0<x<\infty$.}
\label{fig:Dsx}
\end{figure}

The derivative $\frac{\partial}{\partial x}s(x, w)$ for various values of $w$ is shown in figure~\ref{fig:Dsx} which indicates that $\frac{\partial}{\partial x}s(x, w) >0$ except at $x=0$,
and hence $s(x, w)$ is a monotonically increasing function of $x$.
In fact, this can be proved \cite{MS} and is shown in \ref{app:sx_increas}.
Moreover, we have $\frac{\partial}{\partial x}s(0, w)=0$. 
Thus, the following theorem is established:\\

\noindent
{\it Theorem 1.}---

For fixed $w > - \infty $ and any $\bold{k}, \, \bold{k}' \in \frac{2\pi}{L} \mathbb{Z}^3$, 
$  \langle\bold{k}'|u_{\rm BP}^{({\rm rel})}|\bold{k}\rangle \geq 0$,
with the equality holding if and only if $\bold{k} = \bold{k}'=\bold{0}$.\\

Theorem 1 plays a pivotal role in the main result of this paper in Sec.~\ref{sec:PF}.

It should be noted that Theorem 1 is established by using the exact energy eigenvalues and eigenfunctions.
Therefore if some crude approximations are employed,
the above theorem may not be derived correctly.
For example, at the first-order perturbation in $a$, $\langle\bold{k}'|u_{\rm BP}^{({\rm rel})}|\bold{k}\rangle$ is not a non-negative matrix \cite{private}.
To see this, we first calculate
\begin{eqnarray}
\fl
\langle\bold{k}_1', \bold{k}_2'|   \delta^3(\bold{r}_1-\bold{r}_2) \frac{\partial}{\partial r}r  |\bold{k}_1, \bold{k}_2\rangle \nonumber
 =  \frac{1}{V^2} \int_V d^3 \bold{r}_1 \int_V d^3 \bold{r}_2 \,\,
  \rme^{\rmi (\bold{K}-\bold{K}') \cdot \bold{R}} \,
  \rme^{\rmi \bold{k}' \cdot \bold{r}}  \,
  \delta^3(\bold{r}_1-\bold{r}_2) \frac{\partial}{\partial r} \left( r  \rme^{\rmi \bold{k} \cdot \bold{r}}    \right) \nonumber \\
 =  \frac{1}{V} \delta_{\bold{K},\bold{K}'} \, \Lambda (\bold{k}).
 \label{ps-pot-mom-rep}
 \end{eqnarray}
Here $\Lambda (\bold{k})$ is the Tan's $\Lambda$-function,
which is defined as $\Lambda (\bold{k}) =1$ for $k < \infty$,
together with $\int d^3 \bold{k} \, k^{-2} \Lambda (\bold{k}) = 0$ \cite{Tan08-I}.
Clearly, $\Lambda (\bold{k})$ becomes negative for sufficiently large $\bold{k}$.
In fact, the Tan's $\Lambda$-function is written as
 $\Lambda (\bold{k}) := 1-k^{-1} \delta (k^{-1})$ \cite{V12}.
We denote $\langle\bold{k}'|u_{\rm 1st}^{({\rm rel})}|\bold{k}\rangle$ by $\langle\bold{k}'|u_{\rm BP}^{({\rm rel})}|\bold{k}\rangle = \langle\bold{k}'|u_{\rm 1st}^{({\rm rel})}|\bold{k}\rangle + \Or(a^2)$.
Using the general formula discussed in Ref.~\cite{LY58-I} and equation (\ref{ps-pot-mom-rep}),
we obtain
\begin{eqnarray}
\langle\bold{k}_1',\bold{k}_2'| U^{(2)}_{\rm 1st}|\bold{k}_1,\bold{k}_2\rangle \nonumber\\
   =  -  \int_0^{\beta} d \tau \, \rme^{-(\beta-\tau)(k'^2_1+ k'^2_2)/(2m)}
   \langle\bold{k}_1',\bold{k}_2'| v_{\rm ps}(\bold{r}_1-\bold{r}_2) |\bold{k}_1,\bold{k}_2\rangle
   \, \rme^{-\tau(k^2_1+ k^2_2)/(2m)} \nonumber\\
    = \frac{8\pi^3}{V} \delta_{\bold{K},\bold{K}'} \,
     \rme^{- \beta\bold{K}^2/(4m)} 
     \cdot \langle\bold{k}'| u_{\rm 1st}^{({\rm rel})}  |\bold{k}\rangle,
\end{eqnarray}
where
\begin{equation}
  \langle\bold{k}'|u_{\rm 1st}^{({\rm rel})}|\bold{k}\rangle 
  = \frac{(-a)}{2\pi^2 m} \, \frac{\rme^{\beta k'^2/m} - \rme^{\beta k^2/m}}{k'^2-k^2}
     \Lambda (\bold{k}).
\label{u-1st}
\end{equation}
In equation (\ref{u-1st}), the Tan's $\Lambda$-function changes the sign for sufficiently large $k$ and its prefactor cannot change the sign.
Thus, $\langle\bold{k}'|u_{\rm 1st}^{({\rm rel})}|\bold{k}\rangle$ is not a non-negative matrix.

%

\section{Identification of the superfluid transition at the ladder-approximation level}
\label{sec:PF}

\subsection{General description}

\begin{figure}
\begin{picture}(200,85)
\put(80,-6){\includegraphics[width=335pt]{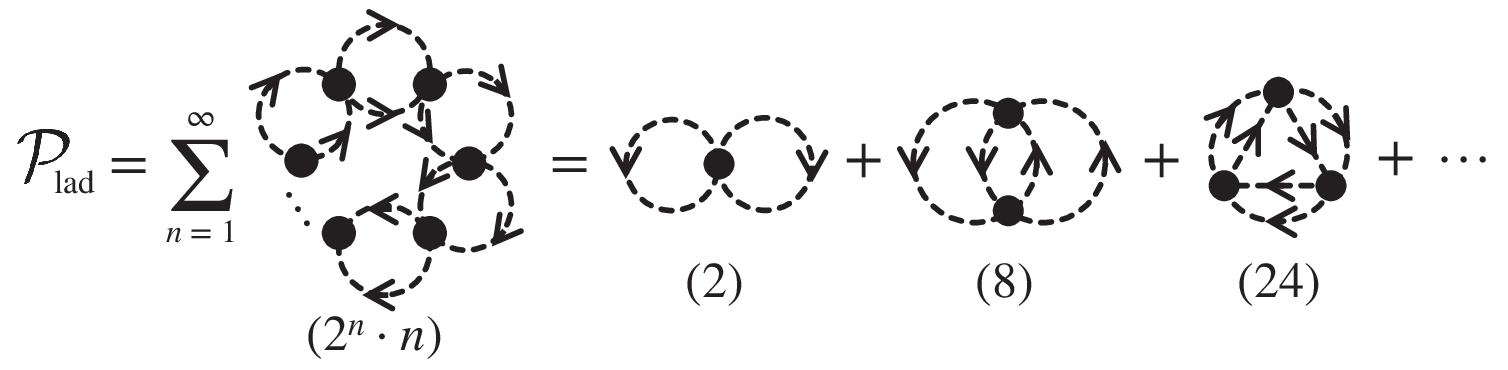}}
\end{picture}
\caption{
$\mathcal{P}_{\rm lad}$ 
as the sum over different ladder-type contracted $0$-graphs.
The corresponding symmetry numbers are shown under the graphs.}
\label{fig:XiLad}
\end{figure}

We consider the sum of the ladder-type contracted $0$-graphs
shown in figure~\ref{fig:XiLad}.
The numerical constant under each term in figure~\ref{fig:XiLad}
is the symmetry number of the corresponding  ladder-type contracted $0$-graph.
The algebraic expression of the sum of the contracted graphs is given as
\begin{eqnarray}
&  \mathcal{P}_{{\rm lad}}
 =  \sum_{n=1}^{\infty} \frac{z^{2n}}{n \cdot 2^n}
 \sum_{\bold{k}_1,\dots,\bold{k}_{2n}}
   \sum_{\sigma_1,\dots,\sigma_{2n}}
     \eta_0 (\bold{k}_1) \dots \eta_0 (\bold{k}_{2n}) \label{eq:XiLad-Irr} \\
& 
\times \langle \bold{k}_1\sigma_1,\bold{k}_2\sigma_2 | \Upsilon_{\rm{A}}^{(2)} | \bold{k}_3\sigma_3,\bold{k}_4\sigma_4 \rangle \cdots 
     \langle \bold{k}_{2n-1}\sigma_{2n-1},\bold{k}_{2n}\sigma_{2n} | \Upsilon_{\rm{A}}^{(2)} | \bold{k}_1\sigma_1,\bold{k}_2\sigma_2\rangle . \nonumber 
\end{eqnarray}
We approximate the grand partition function by the ladder-type contracted $0$-graph as
\begin{equation}
\log \Xi_V (z, \lambda /a) \simeq 
- 2 \frac{V}{\lambda^3} {\rm Li}_{\frac{5}{2}} (-z) 
+ \mathcal{P}_{{\rm lad}},
\label{eq:logX-approximation}
\end{equation}
where ${\rm Li}_l(x):=\sum_{n=1}^\infty x^n/n^{l}$ is the polylogarithm
There is no singularity in ${\rm Li}_{\frac{5}{2}} (-z)$ near the origin and along the positive real axis. 
To examine a phase transition at the ladder-approximation level,
we examine a singularity of $\mathcal{P}_{{\rm lad}}$ on the real positive axis of $z$.
Substituting equation~(\ref{eq:U2_pseudopotential_finiteV}) into equation~(\ref{eq:XiLad-Irr}), we have
\begin{equation}
\mathcal{P}_{{\rm lad}}
 =   \sum_{n=1}^{\infty} \frac{1}{n} \sum_\bold{K}
 \sum_{\bold{p}_1, \dots , \bold{p}_{n}} 
     \mathcal{A}_{\bold{K}} \left( \bold{p}_1, \bold{p}_2  \right)
      \mathcal{A}_{\bold{K}} \left( \bold{p}_2, \bold{p}_3  \right)
     \dots
     \mathcal{A}_{\bold{K}} \left( \bold{p}_n, \bold{p}_1  \right),
\label{eq:ladder-irr-graph2}
\end{equation}
where we define a matrix $\mathcal{A}_{\bold{K}}$, related to 
$U_{\rm BP}^{(2)}$ and $\eta_0 (\bold{k})$, as
\begin{eqnarray}
  \mathcal{A}_{\bold{K}} & \left( \bold{p}, \bold{p}' \right)
  := z^2 \rme^{- \beta \bold{K}^2/(4m)}
     \langle \bold{p} | u_{\rm BP}^{({\rm rel})} | \bold{p}' \rangle  \nonumber \\
& \quad \times  
    \sqrt{ \eta_0 (\bold{K}/2+\bold{p}) \eta_0 (\bold{K}/2-\bold{p})} 
    \sqrt{  \eta_0 (\bold{K}/2+\bold{p}') \eta_0 (\bold{K}/2-\bold{p}')}.
\label{eq:def:MatrixA}
\end{eqnarray}

If $\left| \bold{p} \right|, \left| \bold{p}' \right| \gg \lambda^{-1} \ln z$, $\mathcal{A}_{\bold{K}} \left( \bold{p}, \bold{p}' \right) \simeq 0$.
Then, we can cut off the matrices $\mathcal{A}_{\bold{K}} \left( \bold{p}, \bold{p}' \right) $ and consider a sufficiently large square matrix $\mathcal{A}_{\bold{K}}^{\rm c} \left( \bold{p}, \bold{p}' \right)$ for $0 \leq \left| \bold{p} \right|, \left| \bold{p}' \right| \leq k^{\rm c}$ without any influence on physics by considering of a sufficiently large $k^{\rm c}$.
The matrices $\mathcal{A}_{\bold{K}}^{\rm c} \left( \bold{p}, \bold{p}' \right)$ are clearly symmetric (i.e., $\mathcal{A}_{\bold{K}}^{\rm c} \left( \bold{p}, \bold{p}' \right)
 = \mathcal{A}_{\bold{K}}^{\rm c} \left( \bold{p}', \bold{p}  \right)$)
and primitive (i.e., all the elements are non-negative: $\mathcal{A}_{\bold{K}}^{\rm c} \left( \bold{p}, \bold{p}' \right) \geq 0 $ and its square is positive: $\sum_{\bold{q}} \mathcal{A}_{\bold{K}}^{\rm c} \left( \bold{p}, \bold{q} \right) \mathcal{A}_{\bold{K}}^{\rm c} \left( \bold{q}, \bold{p}' \right) > 0 $).
Therefore, 
the matrices $\mathcal{A}_{\bold{K}}^{\rm c} \left( \bold{p}, \bold{p}' \right)$ satisfy the condition for the Perron-Frobenius theorem:\\

\noindent
{\it{The Perron-Frobenius theorem}}-----

Let $A  = (a_{ij})$ be an $n \times n$  primitive matrix (i.e., $a_{ij} \geq 0$ and $(A^m)_{ij} > 0$ for some natural number $m$ and $1 \leq i,j \leq  n$.).
Then, the following statements hold
\begin{enumerate}
\item[(i)] There is a positive real number $\lambda^{\rm PF}$, called the Perron-Frobenius eigenvalue, such that $\lambda^{\rm PF}$ is an
eigenvalue of $A$ and the absolute value of any other eigenvalue $\lambda$ (possibly, complex) is smaller than $\lambda^{\rm PF}$, $|\lambda| < \lambda^{\rm PF}$.
Thus, the spectral radius is equal to $\lambda^{\rm PF}$.
\item[(ii)] the Perron-Frobenius eigenvalue $\lambda^{\rm PF}$ is a simple root of the characteristic polynomial of $A$.
Consequently, the eigenspace associated with $\lambda^{\rm PF}$ is one-dimensional. 
\item[(iii)] There exists an eigenvector $\bold{u} = (u_1,\dots,u_n)$ of $A$ with eigenvalue $\lambda^{\rm PF}$,
called the Perron-Frobenius vector, such that all components of $\bold{u}$ are positive, i.e., 
$A \bold{u} = \lambda^{\rm PF} \bold{u}$, $u_i > 0$ for $1 \leq i \leq n$.
\item[(iv)] There are no other positive eigenvectors other than $\bold{u}$, i.e. all other eigenvectors must have at least one negative or non-real component.
\item[(v)] Let $B  = (b_{ij})$ be an $n \times n$ primitive matrix and
$\lambda_B^{\rm PF}$ be the Perron-Frobenius eigenvalue of $B$.
If $b_{ij} > a_{ij}$ for $1 \leq i,j \leq  n$,
then $\lambda_B^{\rm PF} > \lambda^{\rm PF}$.
\end{enumerate}
\quad

Now the matrices $\mathcal{A}_{\bold{K}}^{\rm c} \left( \bold{p}, \bold{p}' \right)$ can be diagonalized using orthogonal matrices $U_{\bold{K}} \left( \bold{p}, \bold{p}' \right) $ :
\begin{equation}
\sum_{\bold{p},\bold{p}'}
U_{\bold{K}} \left( \bold{p}, \bold{q} \right) 
\mathcal{A}_{\bold{K}}^{\rm c} \left( \bold{p}, \bold{p}' \right)
U_{\bold{K}} \left( \bold{p}', \bold{q}' \right)
=
\lambda_{\bold{K}}\left( \bold{q} \right) \delta_{\bold{q}, \bold{q}'}.
\label{eq:diagonalize}
\end{equation}
Without loss of generality, we assume that the eigenvalue $\lambda_{\bold{K}} \left( \bold{0} \right)$ is the Perron-Frobenius eigenvalue of the matrix $\mathcal{A}_{\bold{K}}^{\rm c}$ for all $\bold{K}$, that is, if $\bold{q} \not= 0$ then $\lambda_{\bold{K}} \left( \bold{0} \right) 
> \left| \lambda_{\bold{K}} \left( \bold{q} \right)\right|$.
We first establish the following lemma:\\

\noindent
{\it Lemma}---

For any $\bold{q} \not= 0$ and $\bold{K} \not= 0$,
$\lambda_{\bold{0}} ( \bold{0} ) 
> \left| \lambda_{\bold{K}} \left( \bold{q} \right)\right|$.\\

\noindent
{\it Proof.}
Using $\eta_0(\bold{k})
  = \left(1+ z \rme^{-\beta \bold{k}^2/(2m)}\right)^{-1}$,
we have
\begin{eqnarray}
&  \rme^{- \beta \bold{K}^2/(4m)} \,
  \eta_0(\bold{K}/2+\bold{p}) \, \eta_0(\bold{K}/2-\bold{p}) \nonumber \\
&  =\left(\rme^{\beta \bold{K}^2/(8m)}+ z \rme^{-\beta (\bold{p}^2+\bold{K}\cdot\bold{p})/(2m)}\right)^{-1} 
   \left(\rme^{\beta \bold{K}^2/(8m)}+ z \rme^{-\beta (\bold{p}^2-\bold{K}\cdot\bold{p})/(2m)}\right)^{-1} \nonumber \\
&  <\left(1+ z \rme^{-\beta (\bold{p}^2+\bold{K}\cdot\bold{p})/(2m)}\right)^{-1} 
    \left(1+ z \rme^{-\beta (\bold{p}^2-\bold{K}\cdot\bold{p})/(2m)}\right)^{-1} \nonumber \\
&  <\left(1+ z \rme^{-\beta\bold{p}^2/(2m)}\right)^{-2} 
= \eta_0(\bold{p}) \eta_0(-\bold{p}).
\label{eq:lemma-proof1}
\end{eqnarray}
Hence using equation (\ref{eq:def:MatrixA}) and inequality (\ref{eq:lemma-proof1}), 
we obtain
$\mathcal{A}_{\bold{K}}^{\rm c} \left( \bold{p}, \bold{p}' \right) 
< \mathcal{A}_{\bold{0}}^{\rm c} \left( \bold{p}, \bold{p}' \right)$,
for any $\bold{p}, \bold{p}'$.
According to the Perron-Frobenius theorem (v),
we have
$ \lambda_{\bold{K}} \left( \bold{0} \right) < \lambda_{\bold{0}} \left( \bold{0} \right) $.
Combined with 
$ \lambda_{\bold{K}} \left( \bold{0} \right) < \lambda_{\bold{0}} \left( \bold{0} \right) $
and
$\lambda_{\bold{K}} \left( \bold{0} \right) 
> \left| \lambda_{\bold{K}} \left( \bold{q} \right)\right|$, the lemma is proved.
(Q.E.D.)\\

Substituting equation~(\ref{eq:diagonalize}) into equation~(\ref{eq:ladder-irr-graph2}),
we obtain
\begin{equation}
 \mathcal{P}_{{\rm lad}} 
 =  \sum_{n=1}^{\infty} \frac{1}{n} 
 \sum_\bold{K} \sum_{\bold{q}}
   \left( \lambda_{\bold{K}} \left( \bold{q} \right) \right)^n.
\label{eq:ladder-irr-graph3}
\end{equation}
The radius of convergence of a power series $\sum_{n=1}^{\infty} x^n/n$ is $1$.
According to the above lemma,
if $\lambda_{\bold{0}} ( \bold{0})<1$,
the power series (\ref{eq:ladder-irr-graph3}) is convergent with the result
\begin{equation}
 \mathcal{P}_{{\rm lad}} 
 = - \sum_\bold{K} \sum_{\bold{q}} \log \Bigl(  1 -  \lambda_{\bold{K}} \left( \bold{q} \right) \Bigr).
\end{equation}
However, if $\lambda_{\bold{0}} ( \bold{0} )=1$,
then the holomorpy of this function $\mathcal{P}_{{\rm lad}}$ is lost,
indicating that there is a phase transition at $\lambda_{\bold{0}} ( \bold{0} )=1$.
The positive real number $\lambda_{\bold{0}} ( \bold{0} )$ changes with the parameter of the system such as the temperature, scattering length and number density.
Therefore,
the following theorem is established:\\

\noindent
{\it Theorem 2.}---

Let $\lambda_{\bold{0}} ( \bold{0} )$ be the Perron-Frobenius eigenvalue of the primitive matrix 
\begin{equation}
  \mathcal{A}_{\bold{0}}^{\rm c} ( \bold{p}, \bold{p}' )
  = z^2  \langle \bold{p} | u_{\rm BP}^{({\rm rel})} | \bold{p}' \rangle 
  \left(1 - n_{\rm F} (\bold{p})\right)  \left(1 - n_{\rm F} (\bold{p}')\right).
\end{equation} 
Then, at $\lambda_{\bold{0}} ( \bold{0} )=1$,
the holomorpy of the function $\mathcal{P}_{{\rm lad}}$ is lost.\\

According to Theorem 2,
solving
\begin{equation}
\lambda_{\bold{0}} ( \bold{0} )=1,
\end{equation}
we obtain the critical fugacity $z_c = z_c \left(\lambda/a\right)$.
We will give an example of this calculation in the next section.

From the Perron-Frobenius theorem (ii) and Theorem 2,
it is found that there exists a unique fugacity at the phase transition point for any $a$.
Physically this implies that there is no fragmentation of Bose-Einstein condensates of dimers and Cooper pairs at the ladder-approximation level of Lee-Yang contracted $0$-graphs.
Besides, we can show \cite{future} that the thermodynamic potential, which is calculated at the ladder-approximation level of the Lee-Yang contracted $0$-graphs and in the first-order approximation in the coupling constant of $U^{(2)}$,
is identical to the results of Nozieres and Schmitt-Rink (NSR) \cite{NSR85, MRE93}, which include
the effect of the Gaussian fluctuations of Cooper pairs.
Hence, our result indicates that 
if we only consider the range of Gaussian fluctuations of Cooper pairs,
there is no fragmentation in the BCS-BEC crossover.

\subsection{Application to Bose-Einstein condensation of strongly bounded dimers (BEC limit)}

We now apply Theorem 2 to the limit of strongly bounded dimers (BEC limit), i.e., $\frac{\beta}{ma^2} =w^2 \gg 1$.
We first rewrite equation (\ref{eq:U2_pseudopotential_finiteV}) as
\begin{equation}
  \langle\bold{k}'|u_{\rm BP}^{({\rm rel})}|\bold{k}\rangle 
  = 
\langle\bold{k}'|u_{\rm BP}^{({\rm rel, b})}|\bold{k}\rangle +
\frac{\lambda^3}{2^{5/2}\pi^{7/2}} \, \frac{s_{{\rm sc}}(x') -  s_{{\rm sc}}(x)}{x'^2-x^2}  
,
\end{equation}
\begin{equation}
s_{\rm sc}(x)
 =  \frac{1}{x^2+w^2}
   \left(
    w \rme^{-x^2}
     -  \frac{2}{\sqrt{\pi}} \, x F \left(x \right)
     +  \left| w \right| \rme^{w^2} {\rm erfc}  \left( \left| w \right|  \right)
    \right),
\end{equation}

\begin{equation}
\langle\bold{k}'|u_{\rm BP}^{({\rm rel, b})}|\bold{k}\rangle 
= \theta(w) \cdot \rme^{w^2} \psi_{\rm b}(\bold{k}') \psi_{\rm b}^*(\bold{k}) ,
\label{eq:def:urel_b}
\end{equation}
where $\theta \left( x \right) =  \left(  1+ x/ \left| x \right| \right)/2$ and
$\psi_{\rm b}(\bold{k}) = \frac{a^{3/2}}{\pi}\frac{1}{1+ (ka)^2}$.
We can interpret $\psi_{\rm b}(\bold{k})$ as a normalized bound state,
because the Fourier transform of $\psi_{\rm b}(\bold{k})$ is the relative wave function of the bound state $\psi_{\rm b}(\bold{r}) = (2 \pi a)^{-1/2} \, \rme^{-r/a}/r$, which is given in equation (\ref{eq:def:bound-w.f.}).
Keeping the leading order term, we have
\begin{equation}
\langle\bold{k}'|u_{\rm BP}^{({\rm rel})}|\bold{k}\rangle 
  = 
\langle\bold{k}'|u_{\rm BP}^{({\rm rel, b})}|\bold{k}\rangle + O(w^{-1})
= \rme^{w^2} 
\psi_{\rm b}(\bold{p}') \psi_{\rm b}(\bold{p}) + O(w^{-1}).
\end{equation}
Then
\begin{equation}
  \mathcal{A}_{\bold{0}}^{\rm c} ( \bold{p}', \bold{p})
  = z^2 \, \rme^{w^2}
\psi_{\rm b}(\bold{p}') \psi_{\rm b}(\bold{p}) 
                    \left(1 - n_{\rm F} (\bold{p}')\right)\left(1 - n_{\rm F} (\bold{p})\right)+ O(w^{-1}).
\end{equation}
We obtain
\begin{equation}
\fl   \sum_{\bold{q}}  \mathcal{A}_{\bold{0}}^{\rm c} ( \bold{p}, \bold{q})
 \psi_{\rm b}(\bold{q}) \left(1 - n_{\rm F} (\bold{q})\right)
 \simeq z^2 \, \rme^{w^2} 
\sum_{\bold{q}} \biggl[  \psi_{\rm b}(\bold{q}) \left(1 - n_{\rm F} (\bold{q})\right) \biggr]^2
   \psi_{\rm b}(\bold{p}) \left(1 - n_{\rm F} (\bold{p})\right).
\end{equation}
Because of the Perron-Frobenius theorem (iii), (iv) and since $\psi_{\rm b}(\bold{p}) \left(1 - n_{\rm F} (\bold{p})\right) > 0$, 
$\psi_{\rm b}(\bold{p}) \left(1 - n_{\rm F} (\bold{p})\right)$ is the Perron-Frobenius eigenfunction of the matrix $\mathcal{A}_{\bold{0}}^{\rm c}$.
Hence,
$z^2 \, \rme^{w^2} \,
\sum_{\bold{q}} \left[  \psi_{\rm b}(\bold{q}) \left(1 - n_{\rm F} (\bold{p})\right) \right]^2$
is the Perron-Frobenius eigenvalue $\lambda_{\bold{0}} ( \bold{0} )$.
Therefore, at the phase transition point,
\begin{equation}
z_c^2 \, \rme^{w_c^2} \sum_{\bold{q}} 
\left[  \psi_{\rm b}(\bold{q}) \left(1 - n_{\rm F} (\bold{p})\right) \right]^2
=1+ O(w_c^{-1}).
\label{eq:app-to-BEC}
\end{equation}
The effect of the Fermi statistics, which is represented by $\sum_{\bold{q}} 
\left[  \psi_{\rm b}(\bold{q}) \left(1 - n_{\rm F} (\bold{p})\right) \right]^2$, can be evaluated as
\begin{eqnarray}
&\fl \sum_{\bold{q}} 
\left[  \psi_{\rm b}(\bold{q}) \left(1 - n_{\rm F} (\bold{p})\right) \right]^2
 = \frac{a^3}{\pi^2} \int_0^\infty \! dq \, \frac{4\pi q^2}{[1+(qa)^2]^2} 
\left[ 1+z_c\,\rme^{-\beta_c q^2/(2m)}  \right]^{-2} \nonumber\\
& = 1 + \sum_{n=1}^\infty  (1+n) \cdot (-z_c)^n 
\left[
 \left( 1+  n w_c^2 \right)  \rme^{n w_c^2 / 2}\, {\rm erfc} \left( \sqrt{\frac{nw_c^2}{2}}
       \right)
  -  \sqrt{\frac{2n w_c^2}{\pi}} \,
 \right] \nonumber\\
& = 1 +2\sqrt{\frac{2}{\pi}}
\left(  {\rm Li}_{\frac{1}{2}} (-z_c) + {\rm Li}_{\frac{3}{2}} (-z_c) \right) \cdot w_c^{-3} 
+ O(w_c^{-5}).
\label{eq:order-by-order}
\end{eqnarray}

\begin{figure}
\centering
\includegraphics[width=220pt,clip]{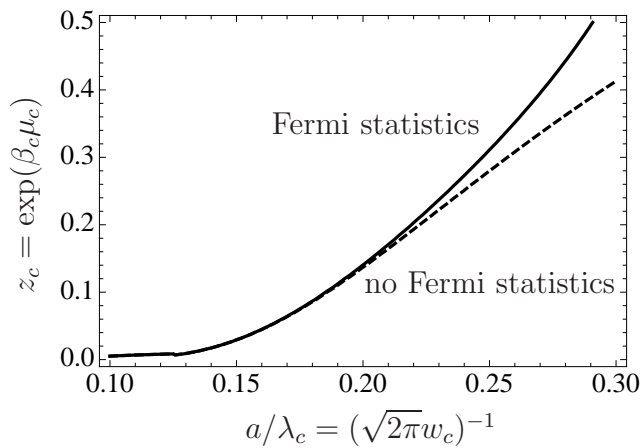}
\put(-150,-15){$a / \lambda_c = (\sqrt{2\pi} w_c)^{-1}$}
\put(-239,40){\rotatebox{90}{$z_c = \exp( \beta_c \mu_c)$}}
\put(-140,100){Fermi statistics}
\put(-105,40){no Fermi statistics}
\caption{Dependence of $z_c = \exp (\beta_c \mu_c )$  on $a/\lambda$ 
at the zeroth approximation (no Fermi statistics, dashed curve) and at the next order (Fermi statistics, solid curve).
The critical fugacity is enhanced by the Fermi statistics.}
\label{fig:BEClimit}
\end{figure}
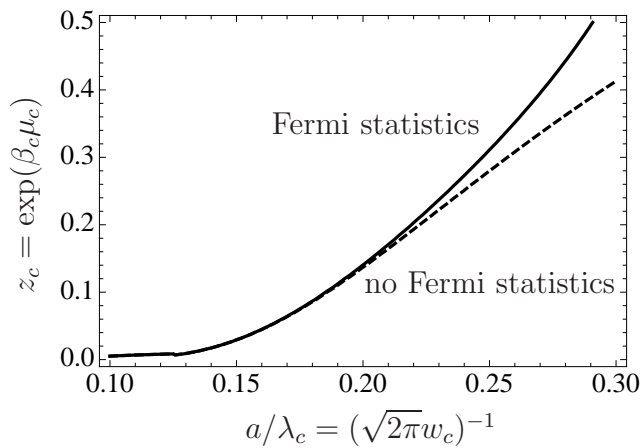

In the zeroth approximation we do not include any $\Or (w_c^{-3})$ terms.
The zeroth-order solution for $z_c$ satisfies $z_c^2 \, \rme^{w_c^2} =1$, and thus we obtain
\begin{eqnarray}
2 \mu_c = -  \frac{1}{ma^2} = - \left| E_{{\rm b}} \right|,
\end{eqnarray}
where $E_{{\rm b}}$ is the binding energy of the dimer.
The zeroth approximation thus leads to a non-interacting-dimer Bose gas.

At the next order, we include the terms which contribute to the order of $w_c^{-3}$. 
Hence, from equations (\ref{eq:app-to-BEC}) and (\ref{eq:order-by-order}) we obtain
\begin{equation}
2 \log z_c + w_c^2 
 +2\sqrt{\frac{2}{\pi}}
\left(  {\rm Li}_{\frac{1}{2}} (-z_c) + {\rm Li}_{\frac{3}{2}} (-z_c) \right) \cdot w_c^{-3} 
+ O(w_c^{-5})
=0.
\end{equation}
Here, ${\rm Li}_{\frac{1}{2}} (-z_c) + {\rm Li}_{\frac{3}{2}} (-z_c)$ is negative.
Hence, the critical fugacity is enhanced by means of the Fermi statistics as shown in figure \ref{fig:BEClimit}.

\section{Summary}

We have applied the Perron-Frobenius theorem to study the superfluid transition of a two-component Fermi gas with a zero-range s-wave interaction.
We have shown that the matrix elements of the two-body cluster functions in momentum representation
are primitive based on the Bethe-Peierls boundary condition or the Fermi pseudopotential.
According to the quantum cluster expansion method of Lee and Yang,
the grand partition function is expressed by the contracted 0-graphs.
We have considered  an infinite series of the ladder-type Lee-Yang contracted 0-graphs and analyzed its singularity.
It has been found that the singularity is governed by the Perron-Frobenius eigenvalue of a certain primitive matrix which is defined in terms of the two-body cluster functions and the Fermi distribution functions.
As a consequence, there exists a unique fugacity at the phase transition point, which indicates that there is no fragmentation of Bose-Einstein condensates of dimers and Cooper pairs at the ladder-approximation level of Lee-Yang contracted $0$-graphs.
An application to the BEC of strongly bounded dimers has also been made.


\ack

We thank S~Endo, T~N~Ikeda, K~Inokuchi, Y~Nishida and T~Ohkuma for useful discussions and Y~Matsuzawa and Y~Suzuki for providing us with the proof of Theorem 1 for the case of $w<0$.
This work was supported by KAKENHI (Nos. 22340114, 22103005, 21540359 and 20102008) and by a Grant-in-Aid for the Global COE Program
``the Next Generation of Physics, Spun from Universality and
Emergence'' from MEXT of Japan. N. K. is partly supported by JSPS
through the ``Funding Program for World-Leading Innovative R\&D on
Science and Technology (FIRST Program)''.

\appendix

\section{Derivation of the two-body cluster function for the s-wave pseudopotential}
\label{app:U2_pseudo}

To calculate the two-body cluster function $U^{(2)}$ in the s-wave approximation, the general formula discussed in Ref.~\cite{LY58-I} is applied.
The formula deals with the case of a central potential and the volume $V=\infty$.
The relationship between the cases of finite $V$ and infinite $V$
is discussed in Ref. \cite{LY60-IV}.
In this section,
we show the subscripts $V$ and $\infty$ to distinguish the cases of finite $V$ and infinite $V$, respectively.

The general formula in Ref.~\cite{LY58-I} needs the set of energy eigenvalues and eigenfunctions,
which are calculated by using the Bethe-Peierls boundary condition (\ref{eq:B-P}) or the Fermi pseudopotential (\ref{eq:def:pseudopotential}).
Here we use the Fermi pseudopotential (\ref{eq:def:pseudopotential}).
The two-body Hamiltonian is
\begin{equation}
   H_{\rm ps}^{(2)}
  = - \frac{1}{2m} \left( \nabla_1^2 + \nabla_2^2 \right)
       + \frac{4\pi a}{m} \delta^3(\bold{r}) \frac{\partial}{\partial r}r.
\end{equation}
Here we introduce the center-of-mass and relative coordinates:
$\bold{R}=(\bold{r}_1+\bold{r}_2)/2$C
$\bold{r}=\bold{r}_1-\bold{r}_2$,
and their absolute values:
$R:=\left|\bold{R}\right|$,
$r:=\left|\bold{r}\right|$.
The Schr\"odinger equation for the relative motion is
\begin{equation}
\left( - \frac{1}{m} \nabla^2    + \frac{4\pi a}{m} \delta^3(\bold{r}) \frac{\partial}{\partial r}r \right) \psi(\bold{r})=
E \psi(\bold{r}).
\end{equation}
The solutions of this equation are continuous scattering states $\psi_{\rm s}(r)$ as shown in equation (\ref{eq:def:scat-w.f.}) with energy $E_{\rm s} = k_{\rm s}^2/m$
and one bound state $\psi_{\rm b}(r)$ as shown in equation (\ref{eq:def:bound-w.f.}) with the binding energy $E_{\rm b} = - 1/(ma^2)$.
The same results can be obtained by the Bethe-Peierls boundary condition.

Using the general formula for the coordinate representation of  the two-body cluster function, we obtain
\begin{equation}
\langle\bold{r}_1'\bold{r}_2'|U^{(2)}_{{\rm BP}, \infty}|\bold{r}_1\bold{r}_2\rangle
 =  \frac{\sqrt{8}}{\lambda^3} \,
 \rme^{- m( \bold{R} - \bold{R}' )^2 / \beta }
      \cdot \langle\bold{r}'| u^{({\rm rel})}_{{\rm BP}, \infty}  |\bold{r}\rangle,
\end{equation}
\begin{eqnarray}
  \langle\bold{r}'|u^{({\rm rel})}_{{\rm BP}, \infty}|\bold{r}\rangle 
 & = \theta (a) \cdot \psi_{\rm b}^*(r)\psi_{\rm b}(r')\rme^{\beta/ma^2}  \nonumber \\  
 &  \,\,\, +\int_0^\infty \!\!\! dk  \rme^{-\beta k^2/m} 
    \biggl[ \psi_{\rm sc}^*(r)\psi_{\rm sc}(r')
  - \frac{\sin (kr) \sin (kr')}{2\pi^2rr'} 
    \biggr],
\end{eqnarray}
where $\theta \left( x \right) =  \left(  1+ x/ \left| x \right| \right)/2$. 
The integration over $k$ gives
\begin{eqnarray}
  \langle\bold{r}'|u^{({\rm rel})}_{{\rm BP}, \infty}|\bold{r}\rangle 
 & =  \frac{1}{4\pi rr'\lambda}
      \Biggl[ \sqrt{2}\,
      \rme^{ -m(r+r')^2/(4\beta)} \nonumber \\
 & \, +   \frac{\lambda}{a} \, \rme^{\beta/(ma^2)}  \rme^{-(r+r')/a} \,
   {\rm erfc} \left( \frac{r+r'}{2}\sqrt{\frac{m}{\beta}} - \frac{1}{a}\sqrt{\frac{\beta}{m}} \right)
 \Biggr],
\end{eqnarray}
where the complementary error function ${\rm erfc} \left(x \right)$ is defined in equation (\ref{eq:def:erfc}).

The momentum representation can be obtained by using
\begin{eqnarray}
 \langle\bold{k}_1'\bold{k}_2'| U^{(2)}_{{\rm BP}, \infty}|\bold{k}_1\bold{k}_2\rangle
  = & \frac{1}{(8\pi^3)^2} \!
   \int \!\! d^3 \bold{r}_1 d^3 \bold{r}'_1 d^3 \bold{r}_2 d^3 \bold{r}'_2 \nonumber\\
 & \times \rme^{- \rmi\sum_{\alpha=1}^2(\bold{k}'_\alpha\cdot\bold{r}'_\alpha -\bold{k}_\alpha\cdot\bold{r}_\alpha)} 
    \langle\bold{r}_1'\bold{r}_2'|U^{(2)}_{{\rm BP}, \infty}|\bold{r}_1\bold{r}_2\rangle ,
\end{eqnarray}
where
$\bold{k}:=(\bold{k}_1-\bold{k}_2)/2$, $\bold{K}:=\bold{k}_1+\bold{k}_2$
and $k:=\left|\bold{k}\right|$.
We write
\begin{equation}
 \langle\bold{k}_1'\bold{k}_2'|U^{(2)}_{{\rm BP}, \infty}|\bold{k}_1\bold{k}_2\rangle
  \equiv \delta^3(\bold{K}-\bold{K}') \, \langle\bold{k}_1'\bold{k}_2'|u_{\rm BP}^{(2)}|\bold{k}_1\bold{k}_2\rangle.
\end{equation}
The function $u^{(2)}$ is defined only for the case with $\bold{k}_1'+\bold{k}_2' = \bold{k}_1+\bold{k}_2$
and is independent of volume.
The cluster function defined for a finite volume $V$ is
\begin{equation}
 \langle\bold{k}_1'\bold{k}_2'|U^{(2)}_{{\rm BP}, V}|\bold{k}_1\bold{k}_2\rangle
 = \frac{8\pi^3}{V} \delta_{\bold{K},\bold{K}'} \,
     \langle\bold{k}_1'\bold{k}_2'|u_{\rm BP}^{(2)}|\bold{k}_1\bold{k}_2\rangle.
\end{equation}

We finally obtain
\begin{equation}
 \langle\bold{k}_1'\bold{k}_2'|u_{\rm BP}^{(2)}|\bold{k}_1\bold{k}_2\rangle
 =  \rme^{- \beta\bold{K}^2/(4m)} \, 
      \langle\bold{k}'| u_{\rm BP}^{({\rm rel})}  |\bold{k}\rangle,
\end{equation}
where
\begin{equation}
  \langle\bold{k}'|u_{\rm BP}^{({\rm rel})}|\bold{k}\rangle 
  = \frac{\lambda}{2^{3/2}\pi^{5/2}} \, \frac{s(k') - s(k)}{k'^2-k^2},
\end{equation}
\begin{eqnarray}
\fl
s(k)
 =  \frac{\sqrt{m}a}{\sqrt{\beta}} \frac{1}{1+(ka)^2}
   \Biggl[
     \rme^{-\beta k^2/m}
     -  \frac{2ka}{\sqrt{\pi}} F \left(\frac{\sqrt{\beta} k}{\sqrt{m}} \right) 
       - \rme^{\beta/(ma^2)}
      {\rm erfc} \left( - \frac{\sqrt{\beta}}{\sqrt{m}a}  \right)
      \Biggr],
\label{eq:gk}
\end{eqnarray}
which gives equation (\ref{eq:def:sx}).

The two-body cluster function $U^{(2)}_{{\rm BP}, \infty}$ for the s-wave pseudopotential has been given in Ref.~\cite{OU06}
for positive $a$,
but the result of this Appendix holds for arbitrary $a$.

\section{Proof of the strictly increasing of the function $s(x,w)$}
\label{app:sx_increas}

In this Appendix, we prove that $\frac{\partial}{\partial x}s(x,w)>0$ for $0<x<\infty$ and $-\infty < w$,
where $\frac{\partial}{\partial x}s(x,w)$ is the derivative of the equation (\ref{eq:def:sx}) which is written as
\begin{eqnarray}
 \frac{\partial}{\partial x} s(x,w)
& =  \frac{2}{\sqrt{\pi}} \frac{x}{(x^2+w^2)^2}
\biggl[ \sqrt{\pi} \left(-w\right)\rme^{-x^2}\left(1+x^2+w^2\right) -1-\frac{1}{x}F\left(x\right) \nonumber \\
& \,\,\, + 2xF\left(x\right)\left(1+x^2+w^2\right) 
-\sqrt{\pi}\frac{\left(-w\right)}{x^2+w^2}\rme^{w^2}{\rm erfc}\left(-w\right)\biggr].
\end{eqnarray}
We will discuss separately the following four cases: 
(i) $-\infty< w \leq -1$, (ii) $-1\leq w\leq-\frac{1}{2}$, (iii) $-\frac{1}{2}<w<0$, (iv) $0\leq w$.

We first organize inequalities which are related to the Dawson's integral and the complementary error function.
The lower and upper bounds for the Dawson's integral are derived, by considering the Taylor series of the integrand of the Dawson's integral and estimating each term, as
\begin{equation}
 F(x) \leq \frac{1}{2x}+\frac{1}{2x^3}+\frac{\rme^{-x^2}}{2x}
\left( \frac{x^2}{2}-2-\frac{1}{x^2}\right) \qquad (0<x) ,
\label{eq:DawsonF-YM1}
\end{equation}
\begin{equation}
\fl
 F(x)\geq \frac{1}{2x}+\frac{1}{4x^3}+\frac{3}{8x^5}+\frac{3\rme^{-x^2}}{8x^5}
\left( -1-\frac{5x^2}{3}-\frac{5x^4}{2}+\frac{5x^6}{6}+\frac{5x^8}{3\cdot 4!}\right)
\quad (0<x),
\label{eq:DawsonF-YM2}
\end{equation}
\begin{equation}
 F\left(x\right) \le \rme^{-x^2}\left(x+\frac{2}{3}x^3\right) \qquad \left(0\le x\le1\right),
\label{eq:DawsonF-YS1}
\end{equation}
\begin{equation}
F\left(x\right) \ge \rme^{-x^2}\left(x+\frac{1}{3}x^3\right).
\label{eq:DawsonF-YS2}
\end{equation}
The upper bounds for the complementary error function are derived as follows: 
\begin{equation}
 {\rm erfc}(x) \leq \frac{1}{\sqrt{\pi}}\frac{\rme^{-x^2}}{x} \qquad (0<x),
\label{eq:Erfc-YM1}
\end{equation}
\begin{equation}
{\rm erfc}(x)\leq \frac{1}{6}\rme^{-x^2}+\frac{1}{2}\rme^{-\frac{4}{3}x^2} \qquad \left(\frac{1}{2} \leq x\right),
\label{eq:Erfc-YM3}
\end{equation}
\begin{equation} 
{\rm erfc}(x)\leq \frac{2}{\sqrt{\pi}}\frac{1}{x+1}\rme^{-x^2} \qquad (0 \leq x),
\label{eq:Erfc-YM2}
\end{equation}
\begin{equation}
{\rm erfc} \left(x\right) \le1-\frac{2}{\sqrt{\pi}}x+\frac{2}{3\sqrt{\pi}}x^3 \qquad \left(0\le x\right).
\label{eq:Erfc-YS}
\end{equation}

\noindent
{\bf (i) $-\infty < w \leq -1 $}

\noindent
The inequalities (\ref{eq:DawsonF-YM1}), (\ref{eq:DawsonF-YM2}) and (\ref{eq:Erfc-YM1}) lead to
\begin{equation}
 \frac{\partial}{\partial x} s(x,w)
 \geq \frac{1}{\sqrt{\pi}}\frac{x \, \rme^{-x^2}}{(x^2+w^2)^2} \tilde{S}_{\rm i}(x,w),
\end{equation}
where
\begin{equation}
\fl
\tilde{S}_{\rm i}(x,w) =
-2-\frac{21x^2}{8}-4w^2-2\sqrt{\pi}(1+x^2+w^2)w
+\frac{5x^4}{4}+\frac{5x^2w^2}{4}.
\end{equation}
For $-\infty < w \leq -1$, we notice that
\begin{equation} 
\frac{{\partial}^2}{{\partial w}^2} \tilde{S}_{\rm i}(x,w) =-12\sqrt{\pi}w-8+\frac{5x^2}{2}>12\sqrt{\pi}-8>0,
\end{equation}
\begin{equation}
\fl 
\frac{\partial}{\partial w}\tilde{S}_{\rm i}(x,-1)=-6\sqrt{\pi}+8-\frac{5x^2}{2}-2\sqrt{\pi}(1+x^2)<-6\sqrt{\pi}+8<0,
\end{equation}
\begin{eqnarray}
\tilde{S}_{\rm i}(x,-1)&=-2-\frac{21}{8}x^2-4+2\sqrt{\pi}(2+x^2)+\frac{5x^4}{4}+\frac{5x^2}{4} \nonumber \\
&= 4\sqrt{\pi}-6+\left( 2\sqrt{\pi}+\frac{5}{2}-\frac{21}{8}\right) x^2>0.
\end{eqnarray}
Thus,  $\tilde{S}_{\rm i}(x,w)>0$.
Therefore, if $-\infty <w\leq -1$, then $\frac{\partial}{\partial x}s(x,w)>0$.\\

\noindent
{\bf (ii) $-1 \leq w \leq -\frac{1}{2} $}

\noindent
The inequalities  (\ref{eq:DawsonF-YM1}), (\ref{eq:DawsonF-YM2}) and (\ref{eq:Erfc-YM3}) lead to
\begin{eqnarray}
 \frac{\partial}{\partial x} s(x,w) &\geq &\frac{1}{\sqrt{\pi}}\frac{\rme^{-x^2}}{x(x^2+w^2)^2}
\tilde{S}_{\rm ii} (x,w),
\end{eqnarray}
where
\begin{eqnarray}
\tilde{S}_{\rm ii} (x,w)=\frac{5x^6}{4}+wh_2(w)x^4
+w\left( h_3(w)-\frac{4}{1-w}-2\sqrt{\pi}h_1(w)\right)x^2  \nonumber \\
\qquad\qquad\,\, +2x^2\left[\left(1+\sqrt{\pi}wh_1(w)\right)\left(\rme^{x^2}-1\right)-\frac{1+w}{1-w}x^2\right],
\label{eq:def:Sii}
\end{eqnarray}
\begin{eqnarray}
h_1(w)&=&\rme^{w^2}{\rm erfc}(-w), \\
h_2(w)&=&\frac{4}{1-w}+\frac{4w}{3}-2\sqrt{\pi}, \\
h_3(w)&=&\frac{4}{1-w}-2\sqrt{\pi}-2\sqrt{\pi}w^2-4w.
\end{eqnarray}
We will evaluate each term in equation (\ref{eq:def:Sii}).
We notice that
$ h'_2(w)=\frac{4}{(1-w)^2}+\frac{4}{3}>0$ and $h_2\left( -\frac{1}{2}\right) =2-2\sqrt{\pi}<0$.
Thus, $h_2(w)<0$ for $w \leq -\frac{1}{2}$.
The inequality (\ref{eq:Erfc-YM3}) lead to $ h_1(w)\leq \frac{1}{6}+\frac{1}{2}\rme^{-\frac{1}{3}w^2}$ for $w\leq -\frac{1}{2}$.
Thus,
\begin{equation}
\fl
 h_3(w)-\frac{4}{1-w}-2\sqrt{\pi} \, h_1(w)
\leq -\frac{5\sqrt{\pi}}{3}-4w-2\sqrt{\pi} \, w^2+\sqrt{\pi} \, \rme^{-\frac{w^2}{3}}
\equiv h_4(w).
\end{equation}
We notice that
$h_4\left(-\frac{1}{2}\right) <0$,
$h_4'\left(-\frac{1}{2}\right) >0$ and
$h_4''(w)=-4\sqrt{\pi}
-\frac{2\sqrt{\pi}}{3}\rme^{-\frac{w^2}{3}}\left(1-\frac{2w^2}{3}\right) <0$
for $-1 \leq w\leq -\frac{1}{2}$.
Thus, $h_4(w)<0$ holds.
Moreover, the following inequality holds:
\begin{equation}
 1+\sqrt{\pi}w \, h_1(w) \geq \frac{1+w}{1-w} \qquad (w<0).
\end{equation}
Hence, $\tilde{S}_{\rm ii} (x,w)>0$.
Therefore, if $-1 \leq w \leq -\frac{1}{2}$, then $\frac{\partial}{\partial x}s(x,w)>0$.\\

\noindent
{\bf (iii) $-\frac{1}{2} < w < 0 $}

\noindent
For $-\frac{1}{2} < w < 0 $, we put $-w=\alpha x\ (\alpha\geq0)$ and discuss separately the following three cases: 
(iii-a) $\frac{1}{2}\leq x$, 
(iii-b) $0<x \leq \frac{1}{2}$ and $1 \leq \alpha$, 
(iii-c) $0<x \leq \frac{1}{2}$ and $0 \leq \alpha \leq1$.\\

\noindent
{\bf (iii-a)} We have $0\leq \alpha \leq 1$.
The inequalities (\ref{eq:DawsonF-YM1}), (\ref{eq:DawsonF-YM2}) and (\ref{eq:Erfc-YM2}) lead to
\begin{equation}
\frac{\partial}{\partial x} s(x,w) \geq \frac{1}{\sqrt{\pi}}\frac{\rme^{-x^2}}{x^5 (1+\alpha^2)^2} \frac{1}{1+\alpha x} \tilde{S}_{\rm iii} (x,\alpha),
\end{equation}
where
\begin{eqnarray}
 \tilde{S}_{\rm iii} & (x, \alpha)
 =  \alpha\left(\frac{1}{4}+\frac{4\alpha^2}{3}\right) x^7
+\left[ \frac{9}{4}+\frac{4\alpha^2}{3}+2\sqrt{\pi}\alpha^2(1+\alpha^2)\right] x^6 \nonumber\\
& -2(2-\sqrt{\pi})\alpha(1+\alpha^2)x^5
   -2(2-\sqrt{\pi})\alpha^2x^4-2(2-\sqrt{\pi})\alpha x^3.
\end{eqnarray}
If $\frac{1}{2} \leq x$ and $0\leq \alpha\leq1$, then
\begin{equation}
\frac{\partial^2}{{\partial x}^2} \tilde{S}_{\rm iii} (x,\alpha) \geq \left[ \frac{135}{16}-22(2-\sqrt{\pi})\alpha\right]x >0,
\end{equation}
\begin{eqnarray}
\frac{\partial}{\partial x} \tilde{S}_{\rm iii} \left(\frac{1}{2}, \alpha \right) =\frac{3\sqrt{\pi}}{8}\alpha^4
+ \left(\frac{5}{8}\sqrt{\pi}-\frac{53}{48}\right) \alpha^3
+\left(\frac{11\sqrt{\pi}}{8}-\frac{7}{4}\right) \alpha^2 \nonumber\\
\qquad\qquad\qquad
 +\left( \frac{17}{8} \sqrt{\pi}-\frac{1081}{256}\right)\alpha+\frac{27}{64}>0,
\label{DSiii1/2}
\end{eqnarray}
\begin{eqnarray}
\tilde{S}_{\rm iii} \left(\frac{1}{2}, \alpha \right) =\frac{\sqrt{\pi}}{32}\alpha^4
+\left(\frac{\sqrt{\pi}}{16}-\frac{11}{96}\right)\alpha^3
+\left(\frac{5\sqrt{\pi}}{32}-\frac{11}{48} \right)\alpha^2 \nonumber \\
\qquad\qquad
+\left(\frac{5\sqrt{\pi}}{16}-\frac{319}{512}\right)\alpha+\frac{9}{256}>0.
\label{Siii1/2}
\end{eqnarray}
Here, inequalities (\ref{DSiii1/2}) and  (\ref{Siii1/2}), in which not all coefficients are positive, follow from evaluating the cubic functions for $0\leq \alpha\leq1$.
Hence, $\tilde{S}_{\rm iii}(x, \alpha) >0$.
Therefore, if $x\geq \frac{1}{2}$ and $-\frac{1}{2} \leq w \leq 0 $, then $\frac{\partial}{\partial x}s(x,w)>0$.\\

\noindent
{\bf (iii-b)} For $-\frac{1}{2} < w < 0 $ and $0<x \leq \frac{1}{2}$,
the inequalities (\ref{eq:DawsonF-YS1}), (\ref{eq:DawsonF-YS2}), (\ref{eq:Erfc-YS}) and
$\rme^{\alpha^2 x^2}\leq 1+\alpha^2 x^2 +\alpha^4 x^4$ (for $0 \leq \alpha x = -w \leq 1$)
lead to
\begin{eqnarray}
\fl
\frac{\partial}{\partial x} s(x,w)
 \geq
 \frac{2}{(1+\alpha^2)^2x^2}
 \Biggl\{ \Bigl[\alpha+\frac{x}{\sqrt{\pi}}(1-\alpha^2)\Bigr] \left(\rme^{-x^2}-1\right)+\alpha x^2\left[(1+\alpha^2)\rme^{-x^2}-\alpha^2\right] \nonumber\\
 \fl \quad
+\frac{2x^3}{3\sqrt{\pi}}\left[\left(3+2\alpha^2\right)\rme^{-x^2}+2\alpha^4\right] 
-\alpha^5x^4 
+\frac{2x^5}{3\sqrt{\pi}}\left[(1+\alpha^2)\rme^{-x^2}+2\alpha^6-\alpha^8x^2\right]
\Biggr\}.
\label{eq:iii-b}
\end{eqnarray}
Now, we consider the case of $1 \leq \alpha$.
Using (\ref{eq:iii-b}), $\rme^{-x^2} - 1\geq - x^2$ and $x \leq 1/(2\alpha)$,
we obtain
\begin{equation}
\fl
\frac{\partial}{\partial x} s(x,w) \geq \frac{4}{3\sqrt{\pi}}\frac{x}{1+\alpha^2} 
\left[
 \left(\frac{3+2 \alpha^2}{1+\alpha^2} \right)\rme^{-1/(4\alpha^2)}-\frac{3\sqrt{\pi}}{4} +\left(2-\frac{3\sqrt{\pi}}{4}\right)\frac{\alpha^4}{1+\alpha^2}
 \right] .
\end{equation}
Since
\begin{equation}
 \left(\frac{3+2\alpha^2}{1+\alpha^2} \right)\rme^{-1/(4\alpha^2)}
 >  \frac{5}{2} \rme^{-1/4}
 >\frac{3\sqrt{\pi}}{4},
 \end{equation}
and $2 > 3\sqrt{\pi}/4$, we obtain $\frac{\partial}{\partial x}s(x,w)>0$.\\

\noindent
{\bf (iii-c)} 
The inequality (\ref{eq:iii-b}) obviously holds.
Using (\ref{eq:iii-b}) and $\rme^{-x^2} \geq 1 - x^2 \geq \frac{3}{4}$,
we obtain
\begin{equation}
\frac{\partial}{\partial x} s(x,w) \geq \frac{1}{\sqrt{\pi}}\frac{x}{(1+\alpha^2)^2} \left[ 4\alpha^2 + 1 -\sqrt{\pi} \alpha (\alpha^2+1) \right].
\end{equation}
If $0\leq \alpha\leq1$, then
$
4\alpha^2 + 1 -\sqrt{\pi} \alpha (\alpha^2+1) > 0.7 > 0
$. 
Thus, $\frac{\partial}{\partial x}s(x,w)>0$.\\

\noindent
{\bf (iv) $0 \leq w$}

\noindent
We divide $\frac{\partial}{\partial x}s(x,w)$ as
$
\frac{\partial}{\partial x}s(x,w) = \frac{\partial}{\partial x}s_1(x,w)+\frac{\partial}{\partial x}s_2(x,w)
$, 
where
\begin{equation}
\frac{\partial}{\partial x} s_1 (x,w)
 =  \frac{2w \, x}{\left(x^2+w^2\right)^2} \rme^{-x^2}
  \left[ \rme^{w^2+x^2} - (1+ w^2 +x^2)   \right],
\end{equation}
\begin{eqnarray}
\frac{\partial}{\partial x} s_2 (x,w)
 =  & \frac{2 \, w^2}{\left(x^2 + w^2\right)^2} 
  \Biggl\{
  \frac{x}{w}  \rme^{w^2} {\rm erf} \left( w  \right)
     - \frac{2}{\sqrt{\pi}} F(x)  \nonumber \\
  &   +   
  \frac{1}{\sqrt{\pi}}\left(1+\frac{x^2}{w^2}\right)
     \left[ (2x^2+1)  F (x)-x \right] \Biggr\}.
\end{eqnarray}
Here
${\rm erf}\left(x \right)
:= \frac{2}{\sqrt{\pi}} \int_{0}^{x} \! dt \, \rme^{-t^2} = 1 - {\rm erfc} (x)$
is the error function.

The following inequalities hold:
\begin{equation}
   \rme^{w^2+x^2} > 1+ w^2 +x^2,
\end{equation}
\begin{eqnarray}
  \frac{\sqrt{\pi}}{2w}  \rme^{w^2} {\rm erf} \left( w  \right)
     >  1
     >  \frac{F(x)}{x},
\end{eqnarray}
\begin{eqnarray}
      (2x^2+1)  F (x)-x >0 \qquad (x>0).
\label{ineq:Dawson integral 1}
\end{eqnarray}
Thus, for $0 \leq w$, 
we obtain $\frac{\partial}{\partial x}s_1(x,w)>0$ and $\frac{\partial}{\partial x}s_2(x,w)>0$.
Then $\frac{\partial}{\partial x}s (x,w)> 0$.


\begin{thebibliography}{99}
\bibitem{MS} The proof of Theorem 1 for $w<0$ in Appendix B is provided by Y~Matsuzawa and Y~Suzuki.
\end{thebibliography}

\end{document}